\documentclass[authoryear,review]{elsarticle}

\usepackage{appendix}
\usepackage{xr}
\usepackage{setspace}
\usepackage[font=small,  margin=2cm]{caption}
\usepackage[tbtags]{amsmath}
\usepackage{amsthm,amssymb,amsfonts}
\usepackage{natbib}
\usepackage{multirow}
\usepackage{lscape}
\journal{Journal of Econometrics \ Templates}
\usepackage{subfigure}
\usepackage{makecell}
\usepackage{exscale}
\usepackage{booktabs}
\usepackage{array}
\usepackage{fullpage}
\usepackage{url}
\usepackage{algorithm}
\usepackage{algpseudocode}
\usepackage{bm}
\usepackage{bbm}
\usepackage{smile}
\usepackage{mathtools}
\usepackage{wrapfig}
\usepackage{lipsum}
\usepackage{mathrsfs}
\usepackage{relsize}
\usepackage{dsfont}
\usepackage{multirow}
\usepackage{apalike}
\usepackage{chngcntr}
\usepackage{graphicx}
\usepackage{placeins}
\usepackage{float}
\usepackage{verbatim}
\usepackage[usenames,dvipsnames,svgnames,table]{xcolor}
\usepackage[colorlinks=true,
linkcolor=blue,
urlcolor=blue,
citecolor=blue]{hyperref}
\usepackage{booktabs} 
\usepackage{siunitx}  

\numberwithin{equation}{section}
\numberwithin{theorem}{section}
\numberwithin{corollary}{section}
\counterwithout{asmp}{section}
\numberwithin{definition}{section}

\makeatletter
\newcommand*{\addFileDependency}[1]{
  \typeout{(#1)}
  \@addtofilelist{#1}
  \IfFileExists{#1}{}{\typeout{No file #1.}}
}
\makeatother

\newcommand*{\myexternaldocument}[1]{%
    \externaldocument{#1}%
    \addFileDependency{#1.tex}%
    \addFileDependency{#1.aux}%
}

\myexternaldocument{supp}

\begin{document}
	
\begin{frontmatter}
\title{Penalized Principal Component Analysis for Large-dimension Factor Model with Group Pursuit}
	
		\author[mythirdaddress]{Yong He\corref{cor1}}
		\author[myfirstaddress]{Dong Liu}
  \author[myfirstaddress]{Guangming Pan}
\author[mythirdaddress]{Yiming Wang}

  \address[mythirdaddress]{Institute for Financial Studies, Shandong University, Jinan, China}
		\address[myfirstaddress]{School of Physical and Mathematical Sciences, Nanyang Technological University, Singapore}

		\begin{abstract}
			
This paper investigates the intrinsic group structures within the framework of large-dimensional approximate factor models, which portrays homogeneous effects of the common factors on the individuals that fall into the same group. To this end, we propose a fusion Penalized Principal Component Analysis (PPCA) method and derive a closed-form solution for the $\ell_2$-norm optimization problem.  We also show the asymptotic properties of our proposed  PPCA estimates. With the PPCA estimates as an initialization, we identify the unknown group structure by a combination of the agglomerative hierarchical clustering algorithm and an information criterion. Then the factor loadings and factor scores are re-estimated conditional on the identified latent groups. Under some regularity conditions, we establish the consistency of the membership estimators as well as that of the group number estimator derived from the information criterion. Theoretically, we show that the post-clustering estimators for the factor loadings with group pursuit  achieve efficiency gains compared to the estimators by traditional PCA method. Thorough numerical studies validate the established theory and a real financial example illustrates the practical usefulness of the proposed method.
		\end{abstract}
	\end{frontmatter}

\section{Introduction}
As a powerful technique for data simplification and dimension reduction, factor models are widely used to extract representative features and explain the generative process of massive variables. They have been  applied in various fields such as financial engineering, economic analysis and biological technology, including studying the expected returns \citep{ross2013arbitrage,ross1977capital,fama1992cross,fama1993common}, risks of portfolios \citep{fan2015risks} and  gene expression data \citep{mayrink2013sparse}. Factor models have also been combined with many modern statistical leaning methods to improve predictive power, such as factor profiled sure independence screening in \cite{wang2012factor}, the estimation of large covariance matrices in \cite{fan2018large} and factor-adjusted multiple testing in \cite{fan2019farmtest}.

 Thus far, the most effective tools for analyzing and predicting large-dimensional time series are the large-dimensional (approximate) factor models, which allow the idiosyncratic errors to be cross-sectionally correlated \citep{NBERw0996}. The estimation and inference for factor
models are crucial in economic studies, particularly in areas such as asset pricing and return forecasting. With the increase of data dimension in these areas, how to accurately estimate and infer the unknown  factor scores and loadings becomes more and more challenging  and attracts growing attention of econometricians. Principal component analysis (PCA) method and Maximum Likelihood Estimation (MLE) method are two common methods for inferring factor models, whose asymptotic properties  have been fully explored in the literature, see for example,  \cite{bai2002determining}, \cite{bai2003inferential},  \cite{forni2005generalized}, \cite{10.1214/11-AOS966}, \cite{bai2016maximum}, \cite{chen2021nonparametric}, \cite{wang2022maximum}, \cite{ma2023group}. In particular, the PCA method, known for
its simplicity and effectiveness,  is closely connected with the factor model and has long been a hot research
topic  in the econometric community.
 An increasing number of work focus on more flexible factor models, for instance, factor models with structural breaks \citep{bates2013consistent,baltagi2017identification}; constrained factor model \citep{tsai2010constrained}; factor model with  latent group structures  \citep{tu2023,he2024large}; robust factor analysis \citep{yu2019robust,Chen2021Quantile,he2022large}; matrix/tesnor factor models \citep{Han2020Tensor,Han2021Rank,chen2022factor,Yu2021Projected,He2023Vector,He2024Matrix,he2024online,Tail-robust}.

In the era of ``big data",  the  cross-sectional dimension in the approximate factor models typically grows extremely large, thus the number of unknown parameters  increases dramatically, which causes great challenges for accurate estimation. To alleviate this issue, \cite{tsai2010constrained} proposed a constrained factor model to study the monthly U.S. excess stock returns where the stocks within the same sector of industry are assumed to share the same loadings on risk factors.  In addition, \cite{xiang2023determining}  divided the cities into four groups and assume the market and government factors have the same impact on house price growth within each group.  Both the studies have assumed the group information is pre-specified, which lack support of economic theory. Consequently, the derived results can be misleading when the group membership is misspecified.  As an alternative, \cite{tu2023} assumed the loading vectors in the factor model have a latent group structure, portraying the homogeneous effects of the common factors on the individuals falling into the same group. \cite{he2024large} also
proposed robust estimators of the factors and factor loadings in the presence of both group structure and heavy-tailedness.  \cite{Shi} proposed to do forecast combination or  portfolio analysis by adopting $\ell_2$-norm relaxation on the weights, deeming that one should assign the same weight to all individual units within the same latent group
whereas potentially distinct weights to the units in different groups. The factor model example for theoretical development in their work also presumes that the factor loadings are directly governed by certain latent group structures.

This paper is inspired by a growing literature on homogeneity in panel data, such as \cite{wang2018homogeneity}, \cite{vogt2017classification}, \cite{MaHuang2017},\cite{chen2019estimating}, \cite{MaHuangZhangLiu+2020}, \cite{chen2021nonparametric}, \cite{guo2022homogeneity}, \cite{wei2023subgroup}, \cite{tu2023} and \cite{wang2024homogeneity}.   Without resorting to any prior information, we develop a completely data-driven unsupervised grouping criterion for the cross-sectional units under the framework of approximate factor model. Our proposed method mainly include the following two steps. In the first step, we propose a fusion Penalized  Principal Component Analysis (PPCA) method to derive an initial estimator with group structure. Interestingly, a closed-form solution for the penalized optimization problem is derived in this paper, which is computationally as efficient as the traditional PCA method. In the second step, we apply the classical hierarchical clustering (AHC) method to the $\ell_1$-distance between the estimated loading vectors by PPCA derived from the first step. We summarize the major contributions of this paper as follows. Firstly, we propose a fusion Penalised PCA (PPCA) method  for estimating the factor scores and factor loadings. A novel $\ell_2$-type fusion penalty is exerted on the pairwise loading vectors to encourage similarity of loadings for individuals within the same group, which interestingly results in a closed-form solution. We study the consistency and asymptotic normality of the PPCA estimators and achieve a faster convergence rate than the traditional PCA estimators by selecting an appropriate tuning parameter. {In the extreme unbalanced case with fixed time horizon $T$ and large cross-sectional dimension $N$, the proposed PPCA estimator remains consistent under mild conditions, whereas the traditional PCA estimator would no longer be consistent.} Secondly,  we introduce a clustering procedure for determining the group membership of each cross-sectional unit and an information criterion for selecting the number of groups.  We investigate the asymptotic properties of the clustering procedure under a relaxed condition on the minimum signal strength and group sizes compared with \cite{tu2023}, which demonstrates that the proposed method can accurately identify the group memberships and estimate the group number with probability approaching to 1. In addition, we have also established the asymptotic properties of the post-grouping  estimators of the factor loadings. Thorough  simulation experiments verify the well finite sample performance of the proposed method in terms of both identifying group memberships and selecting group number. We apply our proposed approach to analyze a portfolio returns dataset, in which we discover intriguing patterns of clustering that contribute to a substantial enhancement in predictive accuracy.

The rest of the paper is structured as follows. Section \ref{sec:1} introduces the approximate factor model with group homogeneity, and we propose the novel PPCA method to estimate the factor scores and factor loadings. {Cross-Validation (CV) criteria are also introduced for the selection of tuning parameters.} Meanwhile this section also introduces the clustering method and an information criterion for identifying the latent group structure and estimating the unknown group number, respectively. Section \ref{sec:3} establishes the asymptotic properties of the proposed PPCA estimators and the subsequent estimators of the clustering procedure. Section \ref{sec:4} displays the finite sample performance of the proposed method by simulation studies, and Section \ref{sec:5} illustrates its practical merits with  an U.S. portfolio return dataset. Section \ref{sec:6} concludes the paper and discusses possible future research directions. All the technical proofs of the theorems and additional empirical results are relegated to the supplementary material.

To end this section, we introduce some notations used throughout this paper.
Denote $\bm{0}_N$ as the $N$-dimensional vector with all elements zero. For a vector $\ba$, we denote $||\ba||_q$ as its $\ell_q$ norm. Denote $\Ib_N$ as the $N$-dimensional identity matrix. For a matrix $\Ab$, $\text{Tr}(\Ab)$ denotes the trace of $\Ab$. We denote $\lambda_j(\Ab)$ as the $j$-th largest eigenvalue of a symmetric matrix $\Ab$ while $\lambda_{\max}(\Ab)$ and $\lambda_{\min}(\Ab)$ correspond to the maximum and minimum eigenvalue, respectively.  We define $||\Ab||_F=\text{Tr}^{1/2}(\Ab^{\top}\Ab)$, $||\Ab||=\lambda^{1/2}_{\max}(\Ab^{\top}\Ab)$ as the Frobenius norm and the spectral norm of $\Ab$ respectively. $X_n\lesssim Y_n$ means there exists a constant $c>0$ such that $X_n\leq cY_n$ for sufficiently large $n$, while $X_n\gtrsim Y_n$ means there exists a constant $c>0$ such that $X_n\geq cY_n$ for sufficiently large $n$, and $X_n\asymp Y_n$ means both $X_n\leq cY_n$ and $X_n\gtrsim Y_n$ hold. The cardinality of a set $\mathcal{G}$ is denoted as $|\mathcal{G}|$. The constants $c$ and $M$ in different lines can be not identical.
\section{Methodology}\label{sec:1}
{In this section, we formally introduce our  method which includes  two main steps. Firstly, we introduce a pairwise fusion penalty  to encourage group pursuit for the factor loadings, thereby putting forward the Penalized PCA method to enhance estimation accuracy for the factor structure. To identify group memberships, we  perform the AHC algorithm to  cluster the individuals into groups, then further deriving refined post-grouping estimators of the factor loadings.}

\subsection{Preliminary}\label{subsec:1}
Consider the static approximate factor model originated from \cite{NBERw0996} for a large panels $\{x_{ti}\}_{t\le T,i\le N}$,
\begin{equation}
	\label{equ:1}
x_{ti}=\sum_{l=1}^{r}b_{il}f_{tl}+e_{ti},\quad\text{for}\quad t=1,\ldots,T, \ \ i=1,\ldots,N,
\end{equation}
where $b_{il}$ are the factor loadings, $f_{tl}$ are the random factor scores,  $e_{ti}$ are the idiosyncratic errors which can be cross-sectionally correlated, $r$ is the number of  common factors. In model (\ref{equ:1}), only $x_{ti}$ can be observed, while $r$, $b_{il}$, $f_{tl}$ and $e_{ti}$ are all unobserved. In this paper, one of our research interests is to refine the estimates of the unknown factor loadings and factor scores by conventional PCA given the prior knowledge of the existence of latent groups of individuals.

The vector and  matrix form of model (\ref{equ:1}) are as follows
\begin{equation}
	\label{equ:2}
\xb_t=\Bb\bm{f}_t+\eb_t,\quad \Xb=\Fb\Bb^{\top}+\Eb
\end{equation}
where $\xb_t^{\top}=(x_{t1},\ldots,x_{tN})$, $\Bb=(\bb_1,\ldots,\bb_N)^{\top}=(b_{il})_{N\times r}$, $\bm{f}_t^{\top}=(f_{t1},\ldots,f_{tr})$, $\eb_t^{\top}=(e_{t1},\ldots,e_{tN})$, $\Xb=(x_{ti})_{T\times N}$, $\Fb=(f_{tl})_{T\times r}$ and $\Eb=(e_{ti})_{T\times N}$. From model (\ref{equ:2}), we can see there exists an identifiable issue among the common factor scores and the factor loadings. In this paper, we assume $\Fb^{\top}\Fb/T=\Ib_r$ to ensure identifiability up to orthogonal transformations, see for example \cite{bai2003inferential} and \cite{10.1214/11-AOS966}  for more detailed discussions on the identifiability issue. Model (\ref{equ:1}) typically can be estimated by Principle Component Analysis (PCA) and Maximum Likelihood Estimation (MLE), see for example \cite{bai2003inferential} and \cite{10.1214/11-AOS966}. In this paper, we primarily focus on the PCA method. A conventional PCA solution to Model (\ref{equ:1}) in \cite{bai2003inferential} is to minimize
$$Q_1(\Bb,\Fb)=(NT)^{-1}||\Xb-\Fb\Bb^{\top}||_F^2,\quad\text{s.t.}\quad\Fb^{\top}\Fb/T=\Ib_r.$$
Let the columns of $\hat{\Fb}$ be $\sqrt{T}$ times the leading $r$ eigenvectors of $\Xb\Xb^{\top}$. Then the solution to minimizing the loss function $Q_1(\Bb,\Fb)$ is $\hat{\Bb}=T^{-1}\Xb^{\top}\hat{\Fb}.$ Therefore, the estimators of the common components are naturally $\hat{c}_{ti}=\sum_{l=1}^{r}\hat{b}_{il}\hat{f}_{tl}$. The theoretical properties of $\hat{b}_{il}$, $\hat{f}_{tl}$, and $\hat{c}_{ti}$ are given in \cite{bai2003inferential} for the large-dimensional case where both $N$ and $T$ go to infinity. Moreover, the factor number $r$ can be estimated by the Information Criterion (IC) by \cite{bai2002determining} or other criteria such as Eigenvalue-Ratio (ER) by \cite{ahn2013eigenvalue}. 

{When it comes to the consistency of the PCA estimators of loadings, as given by \cite{bai2003inferential}, though the convergence rate of the PCA estimator for each loading vector $\bb_i$ enjoys dimensionality blessing, the total number of the loading vectors to be estimated is equal to the number of the cross-sectional units $N$. This would still make the convergence rate of the PCA estimators of loadings  relatively slow. In the extreme cases where $N$ or $T$ is finite, the PCA estimators would no longer be consistent. To improve the convergence rate, it's typically assumed that there exist certain types of structural sparsity, which would further reduce the dimension of the parameters of interest and is quite reasonable  in many real-world scenarios. For example, in psychological research, repetitive testing is essential to ensure reliability, validity, and accuracy in measuring psychological traits. It helps mitigate memory or practice effects, where individuals improve scores due to familiarity rather than actual ability, and allows researchers to track long-term changes in cognition and behavior. When applying factor models to analyze repetitive psychological test scores, due to the homogeneity of test contents and difficulty, a subgroup structure in factor loadings would be quite natural and reasonable.}


Hence, in this paper, we assume the subgroup structures in factor loadings and consider a similar setting as in \cite{Shi}, \cite{tu2023} and \cite{he2024large}. Specifically, we assume the approximate factor model has a latent group homogeneity structure, where factor loadings are the same across units that fall into the same group. Assume there is a partition of index set $\{1,\ldots,N\}$, denoted by $\{\mathcal{G}_1,\ldots,\mathcal{G}_{K_0}\}$ such that
\begin{equation}
	\label{equ:3}
	\bb_{i}=\sum_{k=1}^{K_0}\bb_{\mathcal{G},k}\cdot\mathbbm{1}\{i\in\mathcal{G}_k\}\quad \text{and} \quad \mathcal{G}_k\cap\mathcal{G}_j=\emptyset\quad \text{for} \quad k\neq j,
\end{equation}
where $K_0$ is the true number of latent groups which is assumed to be finite but unknown, $\bb_{\mathcal{G},k}$ is the common value of the loading vectors whose index is in $\mathcal{G}_k$, $\mathbbm{1}\{\cdot\}$ denotes the indicator function and  $\emptyset$ denotes the empty set. From the group homogeneity condition (\ref{equ:3}), it's easy to see that the individuals in the same group share the common loading vectors so as to reduce the number of the loading vectors from $N$ to $K_0$. The group number $K_0$ and the true group $\{\mathcal{G}_1,\ldots,\mathcal{G}_{K_0}\}$ are all unknown, and in the following  section we propose a novel PPCA method to estimate the factor loadings and then apply an AHC procedure to the PPCA estimators to derive the estimators of group memberships.

\subsection{PPCA method for estimating factor loadings and scores }\label{subsec:2}
In model (\ref{equ:1}), only $x_{ti}$ can be observed and our goal in this section is to give a more accurate estimate of the factor scores and factor loadings with prior knowledge of latent group structure. We exploit the pairwise fusion penalty in the literature and propose a Penalized PCA approach with a given $r$ to encourage the sparsity of $(\bb_i-\bb_{j})$, namely,
\begin{equation}\label{equ:opti}
	\rbr{\widehat{\Bb},\widehat{\Fb}}=\argmin_{\Bb,\Fb}\frac{1}{TN}||\Xb-\Fb\Bb^{\top}||_F^2+\tilde{\lambda}\sum_{i<j}||\bb_i-\bb_j||_2^2, \quad \text{s.t.}\quad \Fb^{\top}\Fb/T=\Ib_r,
\end{equation}	
where $\tilde{\lambda}>0$ is a tuning parameter. Clearly, a sufficiently small $\tilde{\lambda}$ implies that there may not exist group structures and our estimator degenerates into the conventional PCA method. On the contrary, a sufficiently large $\tilde{\lambda}$ would force all individuals to share the same factor loadings. Therefore, a moderate $\tilde{\lambda}$ would lead to heterogeneous group structures of the factor loadings. The tuning parameter $\tilde{\lambda}$ typically can be determined by CV method, as will be discussed in Section \ref{sec:selectlam}.

At a first glance, the optimization problem seems to be difficult as there is $N(N-1)$ penalties in (\ref{equ:opti}).  Interestingly, we derive a closed-form solution to the optimization problem.
For better illustration, we first let
\begin{equation}
	\mathcal{L}(\Bb,\Fb)=\frac{1}{TN}||\Xb-\Fb\Bb^{\top}||_F^2+\tilde{\lambda}\sum_{i<j}||\bb_i-\bb_j||_2^2, \quad \text{s.t.}\quad \Fb^{\top}\Fb/T=\Ib_r. \nonumber
\end{equation}
 To minimize $\mathcal{L}(\Bb,\Fb)$, we first define the Laplacian matrix $\bPi=N\Ib_N-\Ab$ where $\Ab$ is a $N\times N$ all one matrix, i.e., $\Ab=(1)_{N\times{N}}$. Meanwhile, we define the normalized form of $\bPi$ as $\bPi_N=\bPi/N$ and write
$$\sum_{i<j}||\bb_i-\bb_j||_2^2=N\text{Tr}(\Bb^{\top}\bPi_N\Bb).$$
Denoting $\lambda=N^2\tilde{\lambda}$, thus we have
\begin{equation}
	\begin{aligned}
		\mathcal{L}(\Bb,\Fb)&=\frac{1}{TN}||\Xb-\Fb\Bb^{\top}||_F^2+\frac{\lambda}{N}\text{Tr}(\Bb^{\top}\bPi_N\Bb)\\
		&=\frac{1}{NT}\text{Tr}(\Xb^{\top}\Xb)-\frac{1}{NT^2}\text{Tr}(\Xb^{\top}\Fb\Fb^{\top}\Xb)+\frac{1}{N}||T^{-1}\Xb^{\top}\Fb-\Bb||_F^2+\frac{\lambda}{N}\text{Tr}(\Bb^{\top}\bPi_N\Bb).
	\end{aligned} \nonumber
\end{equation}
We first assume $\lambda$ and $\hat{\Fb}$ are given while $\hat{\Fb}^{\top}\hat{\Fb}/T=\Ib_r$. Therefore, the solution with respect to the factor loadings would be
\begin{equation}\label{equ:ini_B}
	\hat{\Bb}=\frac{1}{T}\Db^{-1}\Xb^{\top}\hat{\Fb},
\end{equation}
where $\Db=\Ib+\lambda\bPi_N$ and it depends on the tuning parameter $\lambda$ and $N$.
Then plugging $	\hat{\Bb}=\Db^{-1}\Xb^{\top}\hat{\Fb}/T$ back to $\mathcal{L}(\Bb,\Fb)$, we have
$$\mathcal{L}(\hat{\Bb},\Fb)=\frac{1}{NT}\text{Tr}(\Xb^{\top}\Xb)-\frac{1}{NT^2}\text{Tr}(\Fb^{\top}\Xb\Db^{-1}\Xb^{\top}\Fb).$$
Thus the solution with respect to the factor scores ${\Fb}$ can be taken as $\sqrt{T}$ times the leading $r$ eigenvectors of $\Xb\Db^{-1}\Xb^{\top}$. Without loss of generality, we treat $\lambda$ as the tuning parameter hereafter since $\lambda = N^2 \tilde{\lambda}$.
  
{Intuitively, it is straightforward to replace the $\ell_2$ norm penalty by a general $\ell_q$ norm, i.e., $||\bb_i-\bb_j||_q$, which would also be appropriate for group pursuits, as suggested by \cite{MaHuang2017}. However, due to the non-convexity of $\Bb$ and $\Fb$, it is quite challenging to acquire a global minimum up to an orthogonal rotation with $\ell_q$ norm peanlty. This further complicates the analysis of the statistical properties of the global minimizer \citep{wu2016new} and significantly increases computational costs. To inherit the advantage of the PCA estimator while encouraging group structures, we prefer the proposed squared $\ell_2$ norm penalty as in this case we acquire a closed-form solution, i.e, we just need to perform eigen-decomposition of $\Xb\Db^{-1}\Xb^{\top}$, which is computationally as efficient as conventional PCA and significantly alleviates computational burden compared with other $\ell_q$ norms.}

{One disadvantage of the squared $\ell_2$-norm penalty lies in that the estimator $\widehat{\Bb}$ in equation \eqref{equ:ini_B} would not show strict group structure, that is, for some index pairs $(i,j)$, $(\widehat{b}_i-\widehat{b}_j)$ can be sufficiently small but not exactly zero. Therefore, it is necessary to further resort to some clustering algorithms to identify the latent groups. However, it's worth pointing out that our proposed Penalized PCA method would deliver a more efficient estimator in certain cases compared with the conventional PCA method, which benefits the downstream clustering procedure.}

\subsection{Homogeneity pursuit}\label{subsec:3}
As long as we obtain the PPCA estimators in the first step, various clustering algorithms can be used to recover the latent group memberships based on the PPCA estimators, such as the well known $K$-means or Agglomerative hierarchical clustering (AHC).
In this paper, we also adopt the classical AHC algorithm to identify the latent homogeneity structure in the loading vectors with a given factor number $r$, for ease of presentation and comparison with \cite{tu2023} and \cite{he2024large}. In detail, we first define the distance measure between $\bb_i$ and $\bb_j$ as:
\begin{equation}\label{equ:5}
	\Delta_{ij}=\frac{1}{r}||\Hb^{\top}(\bb_i-\bb_j)||_1,
\end{equation}
where $\Hb$ is an $r\times r$ rotation matrix. From (\ref{equ:5}), we can easily see that if $i,j\in \mathcal{G}_k$, $\Delta_{ij}=0$ while $\Delta_{ij}\neq0$ if $i\in\mathcal{G}_{k_1}$ and $j\in\mathcal{G}_{k_2}$, whenever $k_1\neq k_2$. Then denote $\mathbf{\Delta}=(\Delta_{ij})$ as a distance matrix among the loading vectors. To apply the $\text{AHC}$ method, we give a natural estimate of $\Delta_{ij}$ as follows:
\begin{equation}
	\hat{\Delta}_{ij}=\frac{1}{r}||\hat{\bb}_i-\hat{\bb}_j||_1=
	\frac{1}{r}\sum_{l=1}^{r}|\hat{b}_{il}-\hat{b}_{jl}|. \nonumber
\end{equation}
Naturally, denote $\hat{\mathbf{\Delta}}_N$ as the estimator of $\mathbf{\Delta}$, whose elements are define in (\ref{equ:5}).
In the following, we give the $\text{AHC}$ algorithm with a given group number $K$, $1\leq K\leq N$.\\
$\textbf{The AHC algorithm}:$

Step 1. Start with $N$ groups, with each group containing one cross-sectional unit, and set $\hat{\mathbf{\Delta}}=\hat{\mathbf{\Delta}}_N$.

Step 2. Search for the smallest distance among the off-diagonal elements of $\hat{\mathbf{\Delta}}$, and merge the corresponding two groups.

Step 3. Re-calculate the distance among the remaining groups and update the distance matrix $\hat{\mathbf{\Delta}}$. For the distance between any two groups $\mathcal{A}$ and $\mathcal{B}$, we adopt the complete linkage defined as $\mathcal{D}(\mathcal{A},\mathcal{B})=\max_{i\in\mathcal{A},j\in\mathcal{B}}\hat{\Delta}_{ij}$, $\text{i.e.}$, the furthest distance between an individual in $\mathcal{A}$ and an individual in $\mathcal{B}$.

Step 4. Repeat Steps 2-3 until the number of groups reduces to $K$. Denote the current index sets as $\hat{\mathcal{G}}_{1|K}, \hat{\mathcal{G}}_{2|K},\ldots, \hat{\mathcal{G}}_{K|K}$. Then $\hat{\mathcal{G}}(K)=\{\hat{\mathcal{G}}_{1|K}, \hat{\mathcal{G}}_{2|K},\ldots, \hat{\mathcal{G}}_{K|K}\}$ is the estimated group membership for the given group number $K$.
\subsection{Post-grouping estimators and selection of the group number}\label{subsec:4}
This subsection focuses on the selection of group number using an information criterion. 
For a given $K$,  we would obtain the clusters $\hat{\mathcal{G}}(K)=\{\hat{\mathcal{G}}_{1|K}, \hat{\mathcal{G}}_{2|K},\ldots, \hat{\mathcal{G}}_{K|K}\}$ using the $\text{AHC}$ algorithm. To fully make   use of the homogeneity structure, we adopt the same goodness of fit measure as in \cite{tu2023} and \cite{he2024large}. We first minimize the following loss function to obtain the factor loadings using the specific group structure
\begin{equation}
	(NT)^{-1}\sum_{i=1}^{N}\sum_{t=1}^{T}(x_{ti}-\bb_i^\top\hat{\bm{f}}_t)^2, \nonumber
\end{equation}
subject to the group structure condition
\begin{equation}
	\bb_i=\bb_{k|K}\quad \text{for} \quad i\in \hat{\mathcal{G}}_{k|K}, \ \ i=1,\ldots,N,\ \ k=1,\ldots,K, \nonumber
\end{equation}
where $\bb_{1|K},\ldots,\bb_{K|K}$ are the distinct factor loadings in the $K$ groups separately.
With the estimated group structure, we can update the estimates of the factor loadings as follows:
\begin{equation}
	\label{equ:post-estimate}
    \tilde{\bb}_{i|\hat{\mathcal{G}}(K)}=\frac{1}{|\hat{\mathcal{G}}_{k|K}|}(\hat{\Fb}^{\top}\hat{\Fb})^{-1}\hat{\Fb}^{\top}\sum_{i\in\hat{\mathcal{G}}_{k|K}}\underline{\xb}_i=\frac{1}{T|\hat{\mathcal{G}}_{k|K}|}\hat{\Fb}^{\top}\sum_{i\in\hat{\mathcal{G}}_{k|K}}\underline{\xb}_i,
\end{equation}
where $\hat{\Fb}$ is the estimator of the factor scores derived in Section (\ref{subsec:2}), and $\underline{\xb}_i=(x_{1i},\ldots,x_{Ti})^{\top}$. The matrix form of the estimated factor loadings is denoted as follows
$$\tilde{\Bb}_{\hat{\mathcal{G}}(K)}=( \tilde{\bb}_{1|\hat{\mathcal{G}}(K)},\ldots, \tilde{\bb}_{N|\hat{\mathcal{G}}(K)})^{\top}.$$
We then define the following goodness-of-fit measure for the information criterion
\begin{equation}\label{S(K)}
	S(K)=(NT)^{-1}\sum_{i=1}^{N}\sum_{t=1}^{T}\left( x_{ti}-\tilde{\bb}_{i|\hat{\mathcal{G}}(K)}^{\top}\hat{\bm{f}}_t\right) ^2.
\end{equation}
We adopt the following information criterion
\begin{equation}
    \label{equ:IC}
	\mathbb{IC}(K)=\log[S(K)]+K\cdot\rho,
\end{equation}
where $\rho\in(0,1)$ is the tuning parameter depending on $N$ and $T$. Then given a large positive integer $\bar{K}$, we can estimate the group number by minimizing the function $\mathbb{IC}(K)$, denoted as $\hat{K}$, where
\begin{equation}\label{equ:minK}
	\hat{K}=\arg\min_{1\leq K\leq \bar{K}}\mathbb{IC}(K).
\end{equation}

Since given the selected group number $\hat{K}$, we can further obtain the estimated clustering group $\tilde{\mathcal{G}}(\hat{K})=\{\hat{\mathcal{G}}_{1|\hat{K}}, \hat{\mathcal{G}}_{2|\hat{K}},\ldots, \hat{\mathcal{G}}_{\hat{K}|\hat{K}}\}$  by the $\text{AHC}$ algorithm introduced in the last section. Replacing $\hat{\mathcal{G}}(K)$ with $\tilde{\mathcal{G}}(\hat{K})$ in (\ref{equ:post-estimate}), we can get the post-grouping loading matrix $\tilde{\Bb}=\tilde{\Bb}_{\tilde{\mathcal{G}}(\hat{K})}=(\tilde{\bb}_1,\ldots,\tilde{\bb}_N)^{\top}$.      

Furthermore, after obtaining the post-grouping factor loading estimate $\tilde{\Bb}$, we may opt to minimize the following least squares problem to update the estimators of factor scores
\begin{equation}\label{eua:post-factor}
	\min_{\bbf_t}\sum_{i=1}^{N}(x_{ti}-\tilde{\bb}_i^{\top}{\bm{f}}_t)^2,\quad t=1,\ldots,T,
\end{equation}                               
 from which we can easily get the updated estimator $\tilde{\bm{f}}_t=(\tilde{\Bb}^{\top}\tilde{\Bb})^{-1}\tilde{\Bb}^{\top}\xb_t,\quad t=1,\ldots,T,$ and therefore
$\tilde{\Fb}=\Xb\tilde{\Bb}(\tilde{\Bb}^{\top}\tilde{\Bb})^{-1}.$
As an important remark, we can also estimate the factor scores and loadings under the framework of constrained factor models following the work by \cite{tsai2010constrained}, where the constraints for the loading matrix can be specified on the basis of our estimated group structure $\tilde{\mathcal{G}}$.
\subsection{Selection of tuning parameters}\label{sec:selectlam}
In this section, we introduce two CV criteria for the selection of tuning parameters to achieve distinct goals. Given that the PPCA method is contingent upon tuning parameter $\lambda$ and that the estimation accuracy of the PPCA estimators would influence subsequent clustering performance, thus if our primary goal is to achieve  more accurate PPCA estimates for better subsequent clustering outcomes, we can utilize the CV criterion to select a suitable tuning parameter, which minimizes the following objective function at each validation step
\begin{equation}\label{CV1}
||\Xb-\hat{\Fb}(\lambda)\hat{\Bb}(\lambda)^{\top}||_F^2,
\end{equation}
where $\hat{\Bb}(\lambda)$ and $\hat{\Fb}(\lambda)$ are the PPCA estimators  by (\ref{equ:opti}) which correspond to the tuning parameter $\lambda$.

If our goal is to enhance the performance of clustering while concurrently striving to optimize the post-grouping estimators, we can implement an alternative CV criterion. For a given $\lambda\in{\Lambda}$, where ${\Lambda}$ is the candidate set for tuning parameter $\lambda$, we can first obtain the PPCA estimates $\hat{\Fb}(\lambda)$, $\hat{K}(\lambda)$ and $\tilde{\mathcal{G}}(\hat{K}(\lambda))$ using the information criterion (\ref{equ:IC}) and the AHC algorithm, which can be used to obtain the post-grouping estimators for loadings: 
$$\tilde{\bb}_{i|\tilde{\mathcal{G}}(\hat{K}(\lambda))}=\frac{1}{T\left|\tilde{\mathcal{G}}(\hat{K}(\lambda))\right|}\hat{\Fb}(\lambda)^{\top}\sum_{i\in\tilde{\mathcal{G}}(\hat{K}(\lambda))}\underline{\xb}_i.$$
Then we have $\tilde{\Bb}(\lambda)=(\tilde{\bb}_{1|\tilde{\mathcal{G}}(\hat{K}(\lambda))},\ldots,\tilde{\bb}_{N|\tilde{\mathcal{G}}(\hat{K}(\lambda))})^{\top}$ and $\tilde{\Fb}(\lambda)=\Xb\tilde{\Bb}(\lambda)\left(\tilde{\Bb}(\lambda)^{\top}\tilde{\Bb}(\lambda)\right)^{-1}$.
Thus we can further utilize the CV criterion to minimize the following objective function at each validation step
\begin{equation}\label{CV2}
||\Xb-\tilde{\Fb}(\lambda)\tilde{\Bb}(\lambda)^{\top}||_F^2.
\end{equation}
 Although the CV criterion in (\ref{CV2}) entails greater computational burden compared with that in (\ref{CV1}), it would result in more accurate post-grouping estimators. 
 In both simulation studies and real empirical studies, we employ the CV criterion in (\ref{CV2})  to select the tuning parameter $\lambda$.

\section{Theoretical analysis}\label{sec:3}
In this section, we first introduce some mild conditions, which bring the current work into a large-dimensional framework with both serially and cross-sectionally correlated errors. We establish the consistency of the proposed PPCA estimators. We show that the convergence rates depend on the tuning parameters and the loadings. Furthermore, when certain conditions on central limit theorems are satisfied, we derive the asymptotic normalities of the proposed PPCA estimators. We also study the asymptotic properties of the clustering procedure in terms of estimating the group memberships and the unknown group number. We  derive the convergence rates of the post-grouping estimators of the factor loadings.
\subsection{Assumptions}
 First of all, to investigate the asymptotic properties of the PPCA estimators, the following technical assumptions are needed, which are common in the related literature such as \cite{bai2003inferential}, \cite{bai2002determining}, \cite{fan2013large}, \cite{yu2020network} and \cite{he2022large}.
\begin{asmp}\label{asmp:1}
	For all $t\leq{T}$ and $l\leq{r}$, $E(f_{tl})=0$, $E(f_{tl}^2)=1$ and $E(f_{tl}^4)\leq{M}$ for some constant $M>0$. Further assume $\Fb^{\top}\Fb/T=\Ib_r$ almost surely and for any $T$ dimensional vector $\bv$ such that $||\bv||=1$, we have $E||\bv^{\top}\Fb||^2\leq{M}$.
\end{asmp}
\begin{asmp}\label{asmp:2}
	For any $j\leq{N}$ and $l\leq{r}$, assume $|b_{jl}|\leq{M}$. Further assume $N^{-1}\Bb^{\top}\Bb\rightarrow{\bm{\Sigma_B}}$, while the eigenvalues of $\bm{\Sigma_B}$ are distinct and bounded away from zero and infinity, i.e., $c_1\geq{\lambda_1(\bm{\Sigma_B})}>\ldots>\lambda_r({\bm{\Sigma_B}})\geq{c_2}$, for some constants $c_1$, $c_2$$>$0.
\end{asmp}
\begin{asmp}\label{asmp:3}
	The error matrix $\Eb=\Pb_1\mathbf{S}\Pb_2$, where $\mathbf{S}=(\epsilon_{tj})_{T\times{N}}$ with $\epsilon_{tj}$ being independent and $E(\epsilon_{tj})=0$, $E(\epsilon_{tj}^2)=1$, $E(\epsilon_{tj}^4)\leq{M}$, $\Pb_1$ and $\Pb_2$ are two deterministic square matrices. There exists positive constants $c_1$ and $c_2$ so that $c_2\leq{\lambda_t(\Pb_1^{\top}\Pb_1)}\leq{c_1}$ for any $t\leq{T}$ and $c_2\leq{\lambda_j(\Pb_2^{\top}\Pb_2)}\leq{c_1}$ for any $j\leq{N}$. In addition, $\mathbf{S}$ and $\Fb$ are independent.
\end{asmp}

Assumption \ref{asmp:1} requires the latent factors have bounded fourth moments. By assuming that $\Fb^{\top}\Fb=\Ib_r/T$ almost surely and $E||\bv^{\top}\Fb||^2\leq{M}$ for any $||\bv||=1$, we require that the serial dependence among factors can not be too strong and guarantee the model is identifiable up to orthogonal transformations. Assumption \ref{asmp:2} requires the factors are all strong and pervasive and the corresponding eigenvectors are identifiable, which is common in the related literature on factor models. Under Assumption \ref{asmp:3}, time series and cross-sectional dependence in the idiosyncratic errors are allowed such that the factor model is an approximate one rather than a strict one. Similar assumption can be founded in \cite{bai2006determining}, \cite{han2017determining} and \cite{yu2020network}.
\subsection{Convergence rates of PPCA estimators}\label{subsection3.2}
In this section, we establish the consistency of the proposed PPCA estimators. We first give a theoretical result which shows  how the convergence rates depend on the tuning parameter $\lambda$.
\begin{theorem}\label{thm1}
Suppose Assumptions \ref{asmp:1}-\ref{asmp:3} hold, then there exists a sequence of revertible matrices $\Hb$ (depending on $N$,$T$ and the tuning parameter $\lambda$) such that $\Hb^{\top}\Hb\stackrel{p}{\rightarrow}\Ib_r$ as $N,T\rightarrow\infty$, and
\begin{equation}
\begin{aligned}
&\frac{1}{T}||\hat{\Fb}-\Fb\Hb||_F^2=O_p\left( \frac{1}{N}+\frac{\text{Tr}(\Db^{-2})}{{NT^2}}\right) ,\\
&\frac{1}{N}||\hat{\Bb}-\Bb\Hb||_F^2=O_p\left(\frac{1}{N}||(\Db^{-1}-\Ib_N)\Bb||_F^2+\frac{\text{Tr}(\Db^{-2})}{{NT}}+\frac{1}{N^2}\right),\\
&\frac{1}{N}||\hat{\Cb}-\Cb||_F^2=O_p\left(\frac{1}{N}||(\Db^{-1}-\Ib_N)\Bb||_F^2+\frac{\text{Tr}(\Db^{-2})}{{NT}}+\frac{1}{N}\right),
\end{aligned} \nonumber
\end{equation}	
where $\hat{\Cb}=\hat{\Fb}\hat{\Bb}^{\top}$ and $\Cb=\Fb\Bb^{\top}$.
Indeed, we deduce $\Hb=\mathbf{\Omega}(\Fb^{\top}\hat{\Fb}/T)\mathbf{\Lambda}_r^{-1}$ where $\mathbf{\Omega}=\Bb^{\top}\Db^{-1}\Bb/N$,  $\mathbf{\Lambda}_r$ is the diagonal matrix composed of the leading $r$ eigenvalues of $\Xb\Db^{-1}\Xb^{\top}/(NT)$.
\end{theorem}
Theorem \ref{thm1} gives the convergence rates of the estimated factor scores and loadings by the PPCA method. As for the estimator of the factor score matrix, compared with the convergence rate $O_p(N^{-1}+T^{-2})$ in \cite{bai2003inferential}, our PPCA estimator can be more accurate if $1/(1+\lambda)^2= o(1)$ and $T = o(\sqrt{N})$. As for the estimator of the loading matrix,  the  convergence rate is composed of three terms and the first term depends on both the loading's structure and the tuning parameter $\lambda$.  The first term is from the shrinkage bias. We claim that with a suitably selected tuning parameter $\lambda$, the shrinkage bias is always negligible. The second term is a scaling of order $T^{-1}$, while the scaling  depends on the tuning parameter $\lambda$. When $\lambda$ is sufficiently large, the second term would converge to zero. Therefore when $T=o(N^2)$ holds, the PPCA estimator for the loading matrix would be more efficient than that by the conventional PCA method whose  convergence rate is $O_p(N^{-1}+T^{-2})$  by \cite{bai2003inferential}. The convergence rate for the common components is easily derived by combining the convergence rates of the estimators for the loadings and factor scores. {
	Similarly, we consider the special case of unbalanced datasets where $T$ is finite while $N$ tends to infinity. If $1/(1+\lambda)^2= o(1)$, the PPCA estimators of factor scores, factor loadings and common components remain consistent, whereas the PCA estimators by \cite{bai2003inferential} would no longer be consistent}

In the following, we provide an algebraic analysis of the impact of the tuning parameter $\lambda$ on the convergence rates of the estimators for common components. The analysis for factor loadings follows similarly. To derive the best convergence rate, we need to minimize the function $f(\lambda)$, where
\begin{equation}
f(\lambda)=\frac{1}{N}||(\Db^{-1}-\Ib_N)\Bb||_F^2+\frac{1}{(1+\lambda)^2T}+O\left(\frac{1}{NT}\right). \nonumber
\end{equation}
Suppose $\Db^{-1}$ has spectral decomposition $\Db^{-1}=\Ub\bm{\Lambda}_{\Db}\Ub^{\top}$. Then we have
\begin{equation}
f(\lambda)= \frac{{\lambda}^2}{N(1+\lambda)^2}||\Bb^*||_F^2+\frac{1}{T(1+\lambda)^2}+O\left(\frac{1}{NT}\right), \nonumber
\end{equation}
where $\Bb^*=\text{diag}(1,\ldots,1,0)\Ub^{\top}\Bb$, $\text{diag}(1,\ldots,1,0)$ defines a $N\times N$ diagonal matrix whose diagonal elements are $(1,\ldots,1,0)$.
It can be easily deduced that $f(\lambda) $ is minimized by taking $\lambda_1^*=N/(T||\Bb^*||_F^2)$, and we have
$$f(\lambda_1^*)=\frac{(NT)^{-1}||\Bb^*||_F^2}{N^{-1}||\Bb^*||_F^2+T^{-1}}+O\left(\frac{1}{NT}\right),$$
from which we can further see that the optimal convergence rate depends on the size of $N^{-1}||\Bb^*||_F^2$. If $N^{-1}||\Bb^*||_F^2=o(T^{-1})$, the convergence rates for the estimators of loadings and common components are faster than those by conventional PCA method. Even if $N^{-1}||\Bb^*||_F^2\gg T^{-1}$, we would still have $f(\lambda_1^*)\leq cT^{-1}$ for some constant $c$, which guarantees that our PPCA method would not perform worse than the conventional PCA method.

\subsection{Asymptotic normality of the PPCA estimators}
In this section, we aim to derive the asymptotic normality of the proposed PPCA estimators. The following additional assumptions are needed to derive the asymptotic distributions.
\begin{asmp}\label{asmp:4}
	For any $t\leq T$, 
	$$\frac{1}{\sqrt{N}}\Bb^{\top}\Db^{-1}\eb_t\stackrel{d}{\rightarrow}\mathcal{N}(\mathbf{0},\Vb_t),$$ 
	where $\Vb_t=\lim_{N,T\rightarrow\infty}N^{-1}||\pb_{1,t}||^2(\Bb^{\top}\Db^{-1}\Pb_2^{\top}\Pb_2\Db^{-1}\Bb)$, where $\pb_{1,t}$ is the $t$-th row of $\Pb_1$. 
\end{asmp}

\begin{asmp}\label{asmp:5}
	For any $j\leq N$,
	$$\frac{1}{\sqrt{T}}\Fb^{\top}\Eb\frac{\mathbf{d}_j^{-1}}{||\mathbf{d}_j^{-1}||}=\frac{1}{\sqrt{T}}\sum_{t=1}^{T}\bm{f}_t\eb_t^{\top}\frac{\mathbf{d}_j^{-1}}{||\mathbf{d}_j^{-1}||}\stackrel{d}{\rightarrow}\mathcal{N}(\mathbf{0},\Wb_j),$$
	where $\mathbf{d}_j^{-1}$ is the $j$-th column of the matrix $\Db^{-1}$ and $\Wb_j=\lim_{N,T\rightarrow \infty}T^{-1}\text{cov}(\Fb^{\top}\Eb\db_j^{-1}/||\mathbf{d}_j^{-1}||)$.	
\end{asmp}

Assumption \ref{asmp:4} is essential to derive the asymptotic normality for the estimators of factor scores, while Assumption \ref{asmp:5} is exerted to guarantee the asymptotic normality for the estimators of factor loadings. Note that $\Db^{-1}$ contains the tuning parameter $\lambda$ implicitly which may depend on $N$ and $T$. Accordingly, we allow the asymptotic normality to hold with varying $\lambda$. {The assumptions are not stringent as we assume the existence of fourth moments of $\epsilon_{tj}$ in Assumption \ref{asmp:3}. If we further impose additional dependence conditions of $\Eb$ on $\Pb_1$ and $\Pb_2$, the assumptions can be easily verified by the Lyapunov Central Limit Theorem. Furthermore, if $\Pb_1$ and $\Pb_2$ are identity matrix, similar assumptions are also imposed for deriving the limiting distributions of the PCA estimators in \cite{bai2003inferential}.}
\begin{theorem}\label{thm:3}
	Denote the spectral decomposition of $\mathbf{\Omega}$ as $\mathbf{\Omega}=\mathbf{\Gamma}_{\Omega}\mathbf{\Lambda}_{\Omega}\mathbf{\Gamma}_{\Omega}^{\top}$. Assume that Assumptions \ref{asmp:1}-\ref{asmp:4} hold, for the  estimated factor scores, we have \\
	1. if $||\Db^{-1}||_F/\min\{{T,\sqrt{NT}}\}=o(1),$ then we have
	$$\sqrt{N}(\hat{\bm{f}}_t-\Hb^{\top}\bm{f}_t)\stackrel{d}{\rightarrow} \mathcal{N}(\mathbf{0},\mathbf{\Lambda}_{\Omega}^{-1}\mathbf{\Gamma}_{\Omega}^{\top}\Vb_t\mathbf{\Gamma}_{\Omega}\mathbf{\Lambda}_{\Omega}^{-1}),$$
	where $\mathbf{\Omega}=N^{-1}\Bb^{\top}\Db^{-1}\Bb$,  $\Hb$ is the same matrix defined  in Theorem \ref{thm1} and $\Vb_t$ is defined in Assumption \ref{asmp:4}. It can be shown that $\Vb_t=||\pb_{1,t}||^2\lim_{N,T\rightarrow\infty}N^{-1}(\Bb^{\top}\Db^{-1}\Pb_2^{\top}\Pb_2\Db^{-1}\Bb)$, where $\pb_{1,t}$ is the $t$-th row of $\Pb_1$ and the matrices $\Pb_1$ and $\Pb_2$ are defined in Assumption \ref{asmp:3}.\\
	2. otherwise, there exists a constant $c$ such that  $||\Db^{-1}||_F/\min\{T,\sqrt{NT}\}>c,$ then we have
	$$\hat{\bm{f}}_t-\Hb^{\top}\bm{f}_t=O_p\left(||\Db^{-1}||_F/\left(\sqrt{N}\min\{T,\sqrt{NT}\}\right)\right).$$
\end{theorem}
The first part of Theorem \ref{thm:3} shows that when $||\Db^{-1}||_F/\min\{T,\sqrt{NT}\}=o(1)$, the biases of the estimated factor scores are asymptotically negligible. We can also see that the asymptotic variances of the PPCA estimators are asymptotically equivalent to those of the conventional $\text{PCA}$ estimators. The second part show that when $||\Db^{-1}||_F/\min\{T,\sqrt{NT}\}>c$, the convergence rate depends on  $||\Db^{-1}||_F/\left(\sqrt{N}\min\{T,\sqrt{NT}\}\right)$. When $N\gg T$, for the second case, it holds that $\hat{\bm{f}}_t-\Hb^{\top}\bm{f}_t=O_p\left(||\Db^{-1}||_F/(T\sqrt{N})\right)$, which is consistent with the rate  $\text{Tr}(\Db^{-2})/(NT^2)$ derived in Theorem \ref{thm1}. While when $T\gg N$, $\hat{\bm{f}}_t-\Hb^{\top}\bm{f}_t=O_p\left(||\Db^{-1}||_F/(N\sqrt{T})\right)$, which is bounded by $c\sqrt{T}$. In the following Theorem \ref{thm:4}, we establish the asymptotic normality for the PPCA estimators of factor loadings under a scaling condition on the time horizon $T$ and cross-sectional size $N$.
\begin{theorem}\label{thm:4}
	Under Assumptions \ref{asmp:1}-\ref{asmp:3} and Assumption \ref{asmp:5}, with the same notations as in Theorem \ref{thm:3},  for the estimated factor loadings, we have\\
	1. if $T=o(N)$, then we have
	$$\frac{\sqrt{T}(\hat{\bb}_j-\Hb^{\top}\Bb^{\top}\mathbf{d}_j^{-1})}{||\mathbf{d}_j^{-1}||}\stackrel{d}{\rightarrow} \mathcal{N}(\mathbf{0},\mathbf{\Gamma}_{\Omega}^{\top}\Wb_j\mathbf{\Gamma}_{\Omega}),$$
	where $\Wb_j$ is defined in Assumption \ref{asmp:5} and 
$$\Wb_j=\lim_{N,T\rightarrow\infty}\frac{1}{T ||\mathbf{d}_j^{-1}||^2}\left|\left|\Pb_2\mathbf{d}_j^{-1}\right|\right|^2\mathbb{E}(\Fb^{\top}\Pb_1\Pb_1^{\top}\Fb).$$
	2. otherwise, when there exists a constant $c$ such that  $T>cN$, we have
	$$\sqrt{N}(\hat{\bb}_j-\Hb^{\top}\Bb^{\top}\mathbf{d}_j^{-1})=O_p(1).$$
\end{theorem}

\subsection{{Theoretical properties of the AHC clustering procedure and post-grouping estimators}}
In this section, we investigate the asymptotic properties of the clustering procedure and the post-grouping estimators of factor loadings. In the following theorem, we show that with the prior knowledge of the group number $K_0$, the AHC clustering algorithm based on the PPCA estimators would accurately identify the group memberships with probability tending to 1.

\begin{theorem}\label{thm:5}
	Denote $d_n=\min_{1\leq{k_1}\neq{k_2}\leq{K_0}}\min_{i\in{\mathcal{G}_{k_1},j\in{\mathcal{G}_{k_2}}}}||\Hb^{\top}(\bb_i-\bb_j)||$. Suppose Assumptions \ref{asmp:1}-\ref{asmp:3} hold and  $d_n$ satisfies the condition $\max\left\lbrace \sqrt{N/T},N^{-1/2}\right\rbrace =o(d_n)$. If the number of latent group $K_0$ is known a prior and the tunning parameter satisfies $\lambda\asymp\lambda_1^*= N/(T||\Bb^*||_F^2)$, we have
	\begin{equation}
		P\left(\hat{\cG}(K_0)=\cG(K_0)\right) \rightarrow 1, \nonumber
	\end{equation}
	as $N$,$T$$\rightarrow$$\infty$.
\end{theorem}

	In Theorem \ref{thm:5}, we require that the minimum signal strength $d_n$ may diminish towards zero, but at a rate slower than $\max\{\sqrt{N/T},\sqrt{1/N}\}$. Moreover, if $N^{-1}||\Bb^*||_F^2=o(T^{-1})$ and  $N\geq \sqrt{T}$, we allow $d_n$ converges to zero at a more faster rate compared with \cite{tu2023}, which implies that our method is able to handle more challenging clustering problems.	 \cite{chen2019estimating} and \cite{chen2021nonparametric} establish similar consistency results for group pursuit under time-varying coefficient panel data models and functional-coefficient
models, respectively.  However, the estimation error introduced in estimating the unknown factors brings additional technical challenges to the theoretical development for group identification consistency, as also evidenced by \cite{tu2023}.

In practice, the true group number $K_0$ is unknown and in the following, we show that the proposed information criterion   can correctly determine the number of groups asymptotically. To derive the consistency of $\hat{K}$, the following additional assumptions are essential. 

\begin{asmp}\label{asmp:6}
	(a) There exists a constant $\tau_1\in (0,1)$ such that
	\begin{equation}
		\min_{1\leq k\leq K_0}|\mathcal{G}_k|\geq \tau_1\cdot \min\{N,N^{\frac{1}{4}}T^{\frac{1}{2}}\}; \nonumber
	\end{equation}
	(b) The tuning parameter $\rho$  in (\ref{equ:minK}) satisfies 
	(i) $C_{NT}\cdot\rho\rightarrow\infty$,  $C_{NT}=\min\{\sqrt{N_{k_0}},\sqrt{T}\}$, with $N_{K_0}=\min{|\mathcal{G}_k|,\ k=1,\ldots,K_0}$ and (ii) $\rho\rightarrow 0$. 
\end{asmp}

Assumption \ref{asmp:6} (a) exerts a condition on the minimum group size, which relaxes the stringent condition by \cite{tu2023} that  the group sizes is of the same order as $N$.
Assumption \ref{asmp:6} (b) gives some mild conditions on $\rho$ and $C_{NT}$ . For the tuning parameter of information criterion $\rho$, one possible choice that satisfies Assumption \ref{asmp:6} is
$$\rho=\frac{\log(\min\{N_K,T\})}{\min\{N_K,T\}},$$
where $N_K=\min\{|\tilde{\mathcal{G}}_{k|K}|,\ k=1,\ldots,K\}$. This choice is used in our numerical experiments and demonstrates strong empirical performance.
\begin{theorem}\label{thm:7}
	Suppose that Assumptions \ref{asmp:1}-\ref{asmp:6} are satisfied. If the tunning parameter satisfies $\lambda\asymp\lambda_1^*= N/(T||\Bb^*||_F^2)$, we have
	$$P(\hat{K}=K_0)\rightarrow 1.$$
	as $N,T \rightarrow\infty$.
\end{theorem}

This theorem shows that the proposed information criterion can accurately identify the true group numbers. Furthermore, together with Theorem \ref{thm:5}, our proposed method can recover the group structure with probability tending to one, even without prior knowledge of $K_0$. Consequently, to further study the asymptotic behaviors of post-grouping estimators defined in \eqref{equ:post-estimate}, we can directly study the corresponding group-specific oracle estimators $\tilde{\bb}_i^{o}$, which is denoted as
\begin{equation}
	\tilde{\bb}_i^{o}=\frac{1}{T|\mathcal{G}_k|}\hat{\Fb}^{\top}\sum_{i\in\cG_k}\underline{\xb}_{i},\quad \text{for}\quad i\in\mathcal{G}_k,\ \ k=1,\ldots,K_0. \nonumber
\end{equation}

The following theorem gives the convergence rate of the oracle estimator $\tilde{\bb}_i^{o}$.
\begin{theorem}\label{thm:6}
	Suppose that Assumptions \ref{asmp:1}-\ref{asmp:3} are satisfied. Then given the true group structure $\mathcal{G}=\left\lbrace \mathcal{G}_1,\ldots,\mathcal{G}_{K_0}\right\rbrace $, there exists a rotation matrix $\Hb=\mathbf{\Omega}(\Fb^{\top}\hat{\Fb}/T)\mathbf{\Lambda}_r^{-1}$, such that
	$$||\tilde{\bb}_i^{o}-{\Hb}^{\top}\bb_{\mathcal{G},k}||_2^2=O_p\left( \frac{||\Bb^{\top}\Db^{-1}\Bb||_F^2}{TN^2|\mathcal{G}_k|}+\frac{||\Bb^{\top}\Db^{-1}||_F^2}{TN|\mathcal{G}_k|}+\frac{||\Bb^{\top}\Db^{-1}||_F^2}{N^2|\mathcal{G}_k|}+\frac{1}{(1+\lambda)^2T^2}+\frac{1}{NT}\right) ,$$
	for each $i\in \mathcal{G}_k$, $k=1,\ldots,K_0$, where $\Bb$ is the true factor loading matrix and $\Db^{-1}$ is defined in (\ref{equ:ini_B}).
\end{theorem}
In the following we give a more detailed discussion on the convergence rate of the group-specific oracle estimator of factor loadings.  For the first term, we first denote an all one matrix  as $\Ab=(1)_{N\times N}$. Suppose $\Ab$ has the spectral decomposition as $\Ab=\Ub\mathbf{\Lambda}\Ub^{\top}$, where $\Ub$ are composed of the eigenvectors and $\mathbf{\Lambda}$ is a diagonal matrix whose first element on the diagonal is $N$ and the other elements are all zero. By the definition of $\Db^{-1}$, we decompose $\Db^{-1}$ into two parts
$$\Db^{-1}=\frac{\lambda}{(1+\lambda)N}\Ab+\frac{1}{1+\lambda}\Ib_N.$$
Then we can write  $\Bb^{\top}\Db^{-1}\Bb$ as
$$\Bb^{\top}\Db^{-1}\Bb=\frac{\lambda}{(1+\lambda)N}\Bb^{\top}\Ab\Bb+\frac{1}{1+\lambda}\Bb^{\top}\Bb.$$
Therefore the first term is bounded by ${\lambda^2||\Bb^{\top}\Ab\Bb/N||_F^2}/\sbr{(1+\lambda)^2N^2T}+{1}/\sbr{(1+\lambda)^2T}$, that is 
\begin{equation}\label{equ:6}
	\frac{||\Bb^{\top}\Db^{-1}\Bb||_F^2}{TN^2|\mathcal{G}_k|}\lesssim \frac{\lambda^2}{(1+\lambda)^2}\frac{||\Bb^{\top}\Ab\Bb||_F^2}{TN^4|\mathcal{G}_k|}+\frac{1}{(1+\lambda)^2T|\mathcal{G}_k|}.
\end{equation}
By a similar argument, for the second and third term, we have
\begin{equation}\label{equ:7}
	\begin{aligned}
		\frac{||\Bb^{\top}\Db^{-1}||_F^2}{TN|\mathcal{G}_k|}+\frac{||\Bb^{\top}\Db^{-1}||_F^2}{N^2|\mathcal{G}_k|}
		\lesssim \frac{\lambda^2}{(1+\lambda)^2}\frac{||\Bb^{\top}\Ab||_F^2}{N^3|\mathcal{G}_k|}\left(\frac{1}{T}+\frac{1}{N}\right)+\frac{1}{(1+\lambda)^2}\left(\frac{1}{T|\mathcal{G}_k|}+\frac{1}{N|\mathcal{G}_k|}\right).
	\end{aligned}
\end{equation}
Combining (\ref{equ:6}) and (\ref{equ:7}), we have 
\begin{equation}
	||\tilde{\bb}_i^{o}-{\Hb}^{\top}\bb_{\mathcal{G},k}||_2^2\leq O_p\left( \frac{\lambda}{(1+\lambda)^2}\eta(\Bb)+\frac{1}{(1+\lambda)^2}\left(\frac{1}{T|\mathcal{G}_k|}+\frac{1}{N|\mathcal{G}_k|}+\frac{1}{T^2}\right)+\frac{1}{NT}\right) , \nonumber
\end{equation} 
where $\eta(\Bb)=(TN^4|\mathcal{G}_k|)^{-1}||\Bb^{\top}\Ab\Bb||_F^2+\left((TN^3|\mathcal{G}_k|)^{-1}+(N^4|\mathcal{G}_k|)^{-1}\right)||\Bb^{\top}\Ab||_F^2$.
Set
\begin{equation}
	h(\lambda)=\frac{\lambda}{(1+\lambda)^2}\eta(\Bb)+\frac{1}{(1+\lambda)^2}\left(\frac{1}{T|\mathcal{G}_k|}+\frac{1}{N|\mathcal{G}_k|}+\frac{1}{T^2}\right). \nonumber
\end{equation}
Then $h(\lambda)$ would be minimized by taking $\lambda_2^*= \left({(T|\mathcal{G}_k|)}^{-1}+{(N|\mathcal{G}_k|)}^{-1}+T^{-2}\right)/\eta(\Bb)$. Then we can easily deduce that if  $\eta({\Bb})>O\left((T|\mathcal{G}_k|)^{-1}+(N|\mathcal{G}_k|)^{-1}+T^{-2}\right)$,
then $h(\lambda_2^*)=O_p\left({(T|\mathcal{G}_k|)}^{-1}+{(N|\mathcal{G}_k|)}^{-1}+T^{-2}\right)$, and the estimators would still be consistent. Compared with the PPCA estimate of factor loadings, the convergence rate can be improved under the following two cases: (1) $T>N$ and $T=o(N|\mathcal{G}_k|)$; (2) $T\leq N$ and $1/|\mathcal{G}_k|=o(1)$. On the other hand, if $\eta({\Bb})=o\left((T|\mathcal{G}_k|)^{-1}+(N|\mathcal{G}_k|)^{-1}+T^{-2}\right)$, then $h(\lambda_2^*)=\eta({\Bb})$, and compared with the initial  PPCA estimators, the convergence rate for the oracle group-specific estimators of loadings can be further improved by similar discussion. 
Under this framework,  the group-specific estimator with $\lambda_2^*$ would always perform better than the estimator by \cite{tu2023}.

Finally, we study the limiting distributions of the oracle estimators of group-specific factor loadings $\tilde{\bb}_{i}^{o}$. To establish asymptotic normality, we introduce the following assumption:
\begin{asmp}\label{asmp:7}
	For each $k$, as $T,|\mathcal{G}_k|\rightarrow\infty$,
	$$\frac{1}{\sqrt{T|\mathcal{G}_k|}}\sum_{j\in\mathcal{G}_k}\frac{\Bb^{\top}\Db^{-1}\Bb}{N}\sum_{t=1}^{T}\bm{f}_t e_{jt}\stackrel{d}{\rightarrow}\mathcal{N}(\mathbf{0},\mathbf{\Phi}_k),$$
	where
	$$\mathbf{\Phi}_k=\lim_{|\mathcal{G}_k|\rightarrow\infty}\frac{1}{|\mathcal{G}_k|}\sum_{l\in\mathcal{G}_k}\sum_{q\in\mathcal{G}_k}\frac{\Bb^{\top}\Db^{-1}\Bb}{N}\mathbf{\Phi}_{lq}\frac{\Bb^{\top}\Db^{-1}\Bb}{N}, \ \ \mathbf{\Phi}_{lq}=\lim_{T\rightarrow\infty}\frac{1}{T}\sum_{s=1}^{T}\sum_{t=1}^{T}\mathbb{E}[\bm{f}_t\bm{f}_s^{\top}e_{ls}e_{qt}].$$
\end{asmp}

{Assumption \ref{asmp:7} is for the asymptotic normality of the re-estimated factor loadings. Validation of the assumptions is similar to the discussion of Assumption \ref{asmp:4} and \ref{asmp:5} as $\Bb^{\top}\Db^{-1}\Bb/N=O(1)$ and similar assumptions can also be found for deriving the limiting distributions of the PCA estimators in \cite{bai2003inferential}.} 
\begin{theorem}
	\label{thm:10}
	Under Assumptions \ref{asmp:1}-\ref{asmp:7}, as $N,T\rightarrow\infty,$ we have\\
	1. if $|\mathcal{G}_k|=o\left(\min\{N,(1+\lambda)^2T\}\right)$ and $||\Bb^{\top}\Db^{-1}||_F^2(1/N+T/N^2)=o(1)$ then for each $i\in \mathcal{G}_k$, $k=1,\ldots,K_0$,
	$$\sqrt{T|\mathcal{G}_k|}(\tilde{\bb}_i^{o}-{\Hb}^{\top}\bb_{\mathcal{G},k})=\mathbf{\Lambda}_r^{-1}\frac{\hat{\Fb}^{\top}\Fb}{T}\frac{\Bb^{\top}\Db^{-1}\Bb}{N}\frac{1}{\sqrt{T|\mathcal{G}_k|}}\Fb^{\top}\sum_{i\in\mathcal{G}_k}\eb_i+o_p(1)\stackrel{d}{\rightarrow}\mathcal{N}(\mathbf{0},\mathbf{\Lambda}_S^{-1}\mathbf{\Gamma}_S^{\top}\mathbf{\Phi}_k\mathbf{\Gamma}_S\mathbf{\Lambda}_S^{-1}),$$
	where $\mathbf{\Lambda}_r$ is the $r\times r$ diagonal matrix formed by the first $r$ largest eigenvalues of $(NT)^{-1}\Xb\Db^{-1}\Xb$ arranged in the descending order, ${\Hb}$ is given in Theorem \ref{thm1}, and $\mathbf{\Phi}_k$ is defined in Assumption \ref{asmp:7}. \\
	2. otherwise, 
	$$\tilde{\bb}_i^{o}-{\Hb}^{\top}\bb_{\mathcal{G},k}=O_p\left(\frac{1}{\sqrt{NT}}+\frac{1}{T(1+\lambda)}+\frac{||\Bb^{\top}\Db^{-1}||_F}{\sqrt{TN|\mathcal{G}_k|}}+\frac{||\Bb^{\top}\Db^{-1}||_F}{\sqrt{N^2|\mathcal{G}_k|}}\right).$$ 
\end{theorem}

In conclusion, the proposed PPCA estimator achieves a better convergence rate, and under more relaxed conditions, the corresponding AHC method combined with the information criterion can accurately identify the number of groups and recover the group memberships with high probability.   

\section{Simulation studies}\label{sec:4}
In this section, we first conduct numerical simulations to investigate the finite sample performance of our proposed PPCA estimators in comparison with those of the conventional PCA estimators. Then we evaluate the finite sample performance of AHC clustering procedure with PPCA and PCA estimators as initial values respectively.
Throughout the simulations, we assume the number of factors is unknown and apply the information criterion $IC_2$ of \cite{bai2002determining} to determine the factor numbers for PPCA and conventional PCA. For PPCA, the tuning parameter $\lambda$ is selected by the 20-fold CV criterion proposed in Section \ref{sec:selectlam} and the candidate set of $\lambda$ is $\{N,b^{-1}\}$ for $b\in \{0.05,0.1,\ldots,1\}$ with an interval of 0.05. Our simulation results are based on 200 replications.

\subsection{Comparison of the PPCA estimators and the conventional PCA estimators }
In this section, we compare the finite sample performance of PPCA estimators in (\ref{equ:ini_B}) with those of the conventional PCA estimators. We consider the following factor model with two common factors,
$$x_{ti}=b_{i1}f_{t1}+b_{i2}f_{t2}+\sqrt{\theta_i}e_{ti}, i=1,\ldots,N,\quad t=1,\ldots,T,$$
where the factors and the idiosyncratic errors are  generated by the following process: for any $m\leq r$, the  factors are generated by an Auto-Regressive process AR(1), that's, $f_{tm}=0.2f_{t-1,m}+v_{tm}$, where $v_{tm}\stackrel{i.i.d}\sim \mathcal{N}(0,1)$. The idiosyncratic error matrix is generated by $\mathbf{E}=\mathbf{P}_1\mathbf{S}\mathbf{P}_2$, where $\epsilon_{tj}\stackrel{i.i.d}\sim \mathcal{N}(0,\kappa\sigma_e^2)$ and we set $\sigma_e^2=1$ and $\kappa\in\{0.5,0.8,1\}$ for different noise-to-signal ratio. The matrices $\mathbf{P}_1$ and $\mathbf{P}_2$ are both banded matrices with bandwidth equal to 1 and all the non-zero-off-diagonal entries equal to 0.02. We set $\theta_i=4(b_{i1}^2+b_{i2}^2)/3$ for $K_0 = 3$. For $K_0=4$, we set  $\theta_i=b_{i1}^2+b_{i2}^2$.

For a fair comparison, we generate the homogeneous structure similar to that in \cite{tu2023}. Specifically,  we consider the following two scenarios:

\vspace{0.5em}

$\textbf{Scenario 1}$ Set $T\in\{100,150,200\}$, $N\in\{90,120,150\}$ and $K_0=3$ with equal group sizes. For the group structure of factor loadings, we consider the following data generating process

$\cG_1$: $x_{ti}=2f_{t1}+\sqrt{16/3}e_{ti}$;

$\cG_2$: $x_{ti}=2f_{t2}+\sqrt{16/3}e_{ti}$;

$\cG_3$: $x_{ti}=2.4f_{t1}+3.2f_{t2}+\sqrt{64/3}e_{ti}$.

\vspace{0.5em}

$\textbf{Scenario 2}$ Set $T\in\{100,150\}$, $N\in\{120,160,200\}$ and $K_0=4$ with equal group sizes. For the group structure of factor loadings, we consider the following data generating process

$\cG_1$: $x_{ti}=2f_{t1}+\sqrt{4}e_{ti}$;

$\cG_2$: $x_{ti}=2f_{t2}+\sqrt{4}e_{ti}$;

$\cG_3$: $x_{ti}=f_{t1}+3f_{t2}+\sqrt{10}e_{ti}$;

$\cG_4$: $x_{ti}=3f_{t1}+f_{t2}+\sqrt{10}e_{ti}$.

\vspace{0.5em}

 We evaluate the performance of the conventional PCA method and the proposed PPCA method by the empirical Mean Squared Error (MSE) in terms of estimating the common components, defined as  $(NT)^{-1}\sum_{s=1}^{200}||\hat{\Fb}_{(s)}\hat{\Bb}_{(s)}^{\top}-\Fb\Bb^{\top}||_F^2/200$, where $\hat{\Fb}_{(s)}$ and $\hat{\Bb}_{(s)}$ represent the estimators of  $\Fb$ and $\Bb$ at the $s$-th replication.
\begin{table}[htbp]
\caption{MSEs in terms of estimating the common components.}
\renewcommand{\arraystretch}{1.3}
\centering
\label{table1}
\begin{tabular}{ccccccccccc}
	\toprule[2pt]
	\multirow{2}{*}{$T$} & \multirow{2}{*}{$N$} && \multicolumn{2}{c}{$\kappa=0.5$} && \multicolumn{2}{c}{$\kappa=0.8$} && \multicolumn{2}{c}{$\kappa=1$}\\
	\cmidrule(lr){4-5} \cmidrule(lr){7-8} \cmidrule(lr){10-11}
	\multicolumn{2}{c}{}  && PPCA & PCA && PPCA & PCA && PPCA & PCA\\
	\toprule[2pt]
	\multicolumn{11}{c}{Scenario 1}\\
	200&150&&0.1451&0.1483&&0.2325&0.2416&&0.2923&0.3058\\
	200&120&&0.1684&0.1733&&0.2719&0.2829&&0.3419&0.3586\\
	200&90 &&0.2064&0.2116&&0.3353&0.3468&&0.4201&0.4409\\
	150&150&&0.1637&0.1681&&0.2626&0.2746&&0.3276&0.3482\\
	150&120&&0.1883&0.1924&&0.3044&0.3151&&0.3776&0.4004\\
	150&90 &&0.2230&0.2294&&0.3619&0.3776&&0.4581&0.4817\\
	100&150&&0.1988&0.2066&&0.3192&0.3393&&0.4004&0.4318\\
	100&120&&0.2229&0.2293&&0.3581&0.3780&&0.4513&0.4823\\
	100&90 &&0.2622&0.2711&&0.4272&0.4495&&0.5445&0.5763\\
	\hline
	\multicolumn{11}{c}{Scenario 2}\\
	150&200&&0.0923&0.0936&&0.1476&0.1509&&0.1851&0.1895\\
	150&160&&0.1031&0.1050&&0.1648&0.1694&&0.2054&0.2129\\
	150&120&&0.1232&0.1252&&0.1962&0.2024&&0.2429&0.2546\\
	100&200&&0.1159&0.1181&&0.1857&0.1907&&0.2269&0.2397\\
	100&160&&0.1258&0.1294&&0.1996&0.2090&&0.2484&0.2629\\
	100&120&&0.1440&0.1486&&0.2307&0.2406&&0.2881&0.3031\\
	\bottomrule[2pt]
\end{tabular}
\end{table}

Table \ref{table1} presents the MSEs of both PCA and PPCA methods, where the number of factors is all correctly identified (not shown in the tables). Clearly, our proposed PPCA outperforms the PCA estimator under all settings. Moreover, as the sample size gradually increases, the MSEs for both PCA and PPCA estimators  decrease. As the noise level $\kappa$ increases, the MSEs increase accordingly, yet our PPCA estimator remains superior to the PCA estimator. It is noteworthy that as $\kappa$ increases, the advantages of our proposed PPCA estimator in terms of MSEs become more pronounced, further highlighting the stability and superiority of our PPCA estimators, which implies that the PPCA method delivers  more precise estimators, acting as  more competitive initial estimators compared with the PCA method for the subsequent clustering.
\subsection{Clustering performance with PPCA initial estimates}
In this section, we show the clustering performance of the AHC procedure with PPCA and PCA estimators as initial values for both Scenario 1 and Scenario 2. The AHC procedure with PCA estimators as initial values is exactly the clustering method proposed by \cite{tu2023} and we briefly denote it as ``TW".
\begin{table}[htbp]
\caption{Clustering performance and post-grouping estimator accuracy  for Scenario 1.}
\renewcommand{\arraystretch}{1.5}
\centering
\label{table2}
\scalebox{0.8}{
\begin{tabular}{cccccccccccc}
	\toprule[2pt]
	\multirow{2}{*}{$(T,N)$} & \multirow{2}{*}{$\kappa$} & \multirow{2}{*}{Initial Values} & \multicolumn{6}{c}{Clustering-related Indexes} && \multicolumn{2}{c}{Accuracy Indexes}\\
	\cmidrule(lr){4-9} \cmidrule(lr){11-12}
	&  &  &$\hat{K}_{mean}$ & Freq & Rand & aRand & Jaccarrd &Purity && $\mathcal{D}(\tilde{\Bb},\Bb)$ & MSE\\
	\toprule[2pt]
	\multirow{6}{*}{(100,90)} & \multirow{2}{*}{0.5} 
	& PCA &3.000 & $0|0$ & 0.9994 &0.9988 &0.9984 &0.9996  &&0.0166 &0.1562\\
	& &PPCA &3.000 &$0|0$ &1.0000 &1.0000 &1.0000 &1.0000  &&0.0135 &0.1534\\
	& \multirow{2}{*}{0.8} 
	& PCA &3.015 & $0|3$ & 0.9912 &0.9800 &0.9758 &0.9933  &&0.0487 &0.2931\\
	& &PPCA &3.000 & $0|0$ &0.9993 &0.9985 &0.9982 &0.9995  &&0.0200 &0.2493\\
	& \multirow{2}{*}{1}
	&PCA	&3.020	& $3|7$	&0.9785	&0.9525	&0.9450	&0.9817	&&0.0832	&0.4231\\
	& &PPCA	&2.995	& $1|0$	&0.9953	&0.9899	&0.9881	&0.9955	&&0.0355	&0.3301\\
	\hline
	\multirow{6}{*}{(100,120)} & \multirow{2}{*}{0.5} 
	&PCA	&3.000	& $0|0$	&0.9992	&0.9983	&0.9978	&0.9994		&&0.0151	&0.1190\\
	& &PPCA	&3.000	& $0|0$	&1.0000	&1.0000	&1.0000	&1.0000	&&0.0117	&0.1147\\
	& \multirow{2}{*}{0.8} 
	&PCA	&3.020	& $0|4$	&0.9916	&0.9811	&0.9770	&0.9940	&&0.0413	&0.2247\\
	& &PPCA	&3.000	& $0|4$	&0.9996	&0.9991	&0.9988	&0.9997	&&0.0179	&0.1860\\
	& \multirow{2}{*}{1}
	&PCA	&3.020	& $1|5$	&0.9845	&0.9655	&0.9590	&0.9877	&&0.0680	&0.3135\\
	& &PPCA	&3.010	&$0|2$	&0.9961	&0.9913	&0.9892	&0.9972	&&0.0312	&0.2481\\
	\hline
	\multirow{6}{*}{(100,150)} & \multirow{2}{*}{0.5} 
	&PCA	&3.000	&$0|0$	&0.9995	&0.9989	&0.9986	&0.9996	&&0.0138	&0.0942\\
	& &PPCA	&3.000	&$0|0$	&1.0000	&1.0000	&1.0000	&1.0000		&&0.0114	&0.0916\\
	& \multirow{2}{*}{0.8} 
	&PCA	&3.035	&$0|7$	&0.9941	&0.9866	&0.9834	&0.9965		&&0.0335	&0.1716\\
	& &PPCA	&3.005	&$0|1$	&0.9990	&0.9978	&0.9973	&0.9992		&&0.0186	&0.1519\\
	& \multirow{2}{*}{1}
	&PCA	&3.100	&$0|20$	&0.9827	&0.9607	&0.9518	&0.9900	&&0.0701	&0.2570\\
	& &PPCA	&3.005	&$0|1$	&0.9984	&0.9966	&0.9958	&0.9989&&0.0223	&0.1908	\\
	\hline
	\multirow{6}{*}{(150,90)} & \multirow{2}{*}{0.5} 
	&PCA	&3.005	&$0|1$	&0.9996	&0.9991	&0.9989	&0.9998&&0.0140	&0.1542\\
	& &PPCA	&3.000	&$0|0$	&1.0000	&1.0000	&1.0000	&1.0000	&&0.0124	&0.1528\\
	& \multirow{2}{*}{0.8}
	&PCA	&3.000	&$0|0$	&0.9978	&0.9951	&0.9941	&0.9982		&&0.0251	&0.2570\\
	& &PPCA	&3.000	&$0|0$	&0.9997	&0.9994	&0.9993	&0.9998		&&0.0174	&0.2463\\
	& \multirow{2}{*}{1}
	&PCA	&3.005	&$0|1$	&0.9934	&0.9852	&0.9824	&0.9946	&&0.0405	&0.3426\\
	&&PPCA	&3.000	&$0|0$	&0.9995	&0.9990	&0.9986	&0.9996	&&0.0211	&0.3092\\
	\hline
	\multirow{6}{*}{(150,120)} & \multirow{2}{*}{0.5} 
	&PCA	&3.000	&$0|0$	&0.9999	&0.9998	&0.9998	&0.9999		&&0.0107	&0.1147\\
	& &PPCA	&3.000	&$0|0$	&1.0000	&1.0000	&1.0000	&1.0000		&&0.0101	&0.1142\\
	& \multirow{2}{*}{0.8} 
	&PCA	&3.000	&$0|0$	&0.9994	&0.9986	&0.9982&	0.9995		&&0.0170	&0.1870\\
	&&PPCA	&3.000	&$0|0$	&0.9998	&0.9996	&0.9995	&0.9998	&&0.0142 &0.1841\\
	& \multirow{2}{*}{1}
	&PCA	&3.005	&$0|1$	&0.9965	&0.9922	&0.9904	&0.9974		&&0.0272	&0.2460\\
	&&PPCA	&3.000	&$0|0$	&0.9996	&0.9992	&0.9990	&0.9997		&&0.0169	&0.2312\\	
	\hline
	\multirow{6}{*}{(150,150)} 
	& \multirow{2}{*}{0.5} &PCA	&3.000	&$0|0$	&0.9998	&0.9997	&0.9996	&0.9999		&&0.0094	&0.0923\\
	&&PPCA	&3.000	&$0|0$	&1.0000	&1.0000	&1.0000	&1.0000		&&0.0086	&0.0915	\\
	& \multirow{2}{*}{0.8} 
	&PCA	&3.010	&$0|2$	&0.9992	&0.9982	&0.9977	&0.9998	&&0.0142	&0.1489\\
	&&PPCA	&3.005	&$0|1$	&0.9999	&0.9999	&0.9998	&1.0000		&&0.0115	&0.1468\\
	& \multirow{2}{*}{1}
	&PCA	&3.015	&$0|3$	&0.9957	&0.9903	&0.9882	&0.9970	&&0.0263	&0.2029\\
	&&PPCA	&3.005	&$0|1$	&0.9998	&0.9997	&0.9996	&0.9999	&&0.0137	&0.1841\\
	\bottomrule[2pt]
\end{tabular}
}
\end{table}

\begin{table}[htbp]
\caption{Clustering performance and post-grouping estimator accuracy  for Scenario 2.}
\renewcommand{\arraystretch}{1.5}
\centering
\label{table3}
\scalebox{0.8}{
\begin{tabular}{cccccccccccc}
	\toprule[2pt]
		\multirow{2}{*}{$(T,N)$} & \multirow{2}{*}{$\kappa$} & \multirow{2}{*}{Initial Values} & \multicolumn{6}{c}{Clustering-related Indexes} && \multicolumn{2}{c}{Accuracy Indexes}\\
	\cmidrule(lr){4-9} \cmidrule(lr){11-12}
	&  &  &$\hat{K}_{mean}$ & Freq & Rand & aRand & Jaccarrd &Purity && $\mathcal{D}(\tilde{\Bb},\Bb)$ & MSE\\
	\toprule[2pt]
	\multirow{6}{*}{(100,120)} & \multirow{2}{*}{0.5} 
	&PCA	&4.000	&$3|3$	&0.9960	&0.9899	&0.9866	&0.9944		&&0.0210	&0.0853\\
	&&PPCA	&3.995	&$1|0$	&0.9981	&0.9953	&0.9936	&0.9975	&&0.0174	&0.0818\\
	& \multirow{2}{*}{0.8}
	&PCA	&3.865	&$20|1$	&0.9749	&0.9425	&0.9323	&0.9566	&&0.0501	&0.1700\\
	&&PPCA	&3.905	&$14|0$	&0.9827	&0.9600	&0.9519	&0.9706	&&0.0421	&0.1558\\
	& \multirow{2}{*}{1}
	&PCA	&3.540	&$61|2$	&0.9316	&0.8513	&0.8347	&0.8729	&&0.0922	&0.2837\\
	&&PPCA	&3.690	&$41|0$	&0.9521	&0.8945	&0.8807	&0.9132	&&0.0737	&0.2420\\
	\hline
	\multirow{6}{*}{(100,160)} & \multirow{2}{*}{0.5} 
	&PCA	&4.000	&$0|0$	&0.9988	&0.9969	&0.9957	&0.9987		&&0.0150	&0.0602\\
	&&PPCA	&4.000	&$0|0$	&0.9994	&0.9984	&0.9976	&0.9994		&&0.0142	&0.0593\\
	& \multirow{2}{*}{0.8} 
	&PCA	&3.845	&$22|3$	&0.9724	&0.9389	&0.9302	&0.9510	&&0.0486	&0.1432\\
	&&PPCA	&3.905	&$13|1$	&0.9809	&0.9564	&0.9483	&0.9679		&&0.0414	&0.1265\\
	& \multirow{2}{*}{1}
	&PCA	&3.315	&$81|1$	&0.9089	&0.8101	&0.7992	&0.8232	&&0.1043	&0.2876\\
	&&PPCA	&3.495	&$60|0$	&0.9299	&0.8516	&0.8396	&0.8668	&&0.0885	&0.2423\\
	\hline
	\multirow{6}{*}{(100,200)} & \multirow{2}{*}{0.5} 
	&PCA	&4.015	&$3|4$	&0.9983	&0.9954	&0.9934	&0.9986		&&0.0153	&0.0495\\
	&&PPCA	&4.000	&$0|0$	&0.9990	&0.9975	&0.9964	&0.9990		&&0.0138	&0.0486\\
	& \multirow{2}{*}{0.8} 
	&PCA	&3.950	&$12|4$	&0.9829	&0.9583	&0.9471	&0.9740		&&0.0417	&0.1073\\
	&&PPCA	&3.990	&$2|1$	&0.9907	&0.9763	&0.9678	&0.9884	&&0.0327	&0.0926\\	
	& \multirow{2}{*}{1}
	&PCA	&3.920	&$20|10$	&0.9694	&0.9267	&0.9089	&0.9532		&&0.0619	&0.1524\\
	&&PPCA	&3.985	&$4|2$	&0.9846	&0.9604	&0.9466	&0.9810	&&0.0458	&0.1233\\
	\hline
	\multirow{6}{*}{(150,120)} & \multirow{2}{*}{0.5} 
	&PCA	&4.000	&$0|0$	&0.9998	&0.9995	&0.9993	&0.9998	&&0.0124	&0.0773\\
	&&PPCA	&4.000	&$0|0$	&0.9998	&0.9996	&0.9995	&0.9998	&&0.0122	&0.0772\\
	& \multirow{2}{*}{0.8} 
	&PCA	&3.955	&$5|0$	&0.9927	&0.9841	&0.9822	&0.9869	&&0.0246	&0.1377\\
	&&PPCA	&3.965	&$4|0$	&0.9950	&0.9895	&0.9885	&0.9907	&&0.0216	&0.1334\\
	& \multirow{2}{*}{1}
	&PCA	&3.895	&$13|1$	&0.9837	&0.9649	&0.9612	&0.9700		&&0.0364	&0.1858\\
	&&PPCA	&3.940	&$7|0$	&0.9905	&0.9793	&0.9765	&0.9830	&&0.0301	&0.1731\\
	\hline
	\multirow{6}{*}{(150,160)} & \multirow{2}{*}{0.5} 
	&PCA	&4.005	&$0|1$	&0.9998	&0.9995	&0.9993	&0.9999		&&0.0104	&0.0593\\
	&&PPCA	&4.000	&$0|0$	&0.9999	&0.9997	&0.9996	&0.9999		&&0.0103	&0.0592\\
	& \multirow{2}{*}{0.8} 
	&PCA	&3.980	&$3|1$	&0.9955	&0.9901	&0.9885	&0.9924		&&0.0193	&0.1019\\
	&&PPCA	&3.975	&$3|0$	&0.9960	&0.9912	&0.9900	&0.9928	&&0.0180	&0.1012\\
	& \multirow{2}{*}{1}
	&PCA	&3.860	&$22|1$	&0.9793	&0.9541	&0.9475	&0.9613	&&0.0393	&0.1548\\
	&&PPCA	&3.895	&$16|0$	&0.9838	&0.9637	&0.9582	&0.9705	&&0.0346	&0.1472\\
	\hline
	\multirow{6}{*}{(150,200)} & \multirow{2}{*}{0.5} 
	&PCA	&4.000	&$0|0$	&0.9994	&0.9986	&0.9981	&0.9994	&&0.0102	&0.0482\\
	&&PPCA	&4.000	&$0|0$	&0.9998	&0.9996	&0.9994	&0.9998	&&0.0096	&0.0476\\
	& \multirow{2}{*}{0.8} 
	&PCA	&3.980	&$2|0$	&0.9965	&0.9923	&0.9911	&0.9940	&&0.0173	&0.0818\\
	&&PPCA	&4.000	&$0|0$	&0.9992	&0.9980	&0.9971	&0.9992	&&0.0143	&0.0771\\
	& \multirow{2}{*}{1}
	&PCA	&3.990	&$3|2$	&0.9939	&0.9851	&0.9805	&0.9914	&&0.0256	&0.1063\\
	&&PPCA	&3.995	&$1|0$	&0.9959	&0.9894	&0.9854	&0.9950&&0.0228	&0.1026\\
	\bottomrule[2pt]
\end{tabular}
}
\end{table}

To assess the accuracy with PPCA and PCA initial values, we report the average MSE of post-group estimator of common components. Furthermore, we also evaluate the distance of the estimated loading space $\text{span}(\tilde{\Bb})$ and the true loading space $\text{span}(\Bb)$. In view of the identifiability issue, we apply the subspace metrics used in \cite{Yu2021Projected} to evaluate the empirical performance, that is,
$$\mathcal{D}(\tilde{\Bb},\Bb)=\left(1-\frac{1}{r}\text{Tr}(\tilde{\Qb}\tilde{\Qb}^{\top}\Qb\Qb^{\top})\right)^{1/2},$$
where $\Qb$ and $\tilde{\Qb}$ are the left singular-vector matrices of the true loading $\Bb$ and its estimator $\tilde{\Bb}$. Clearly, $\mathcal{D}(\tilde{\Bb},\Bb)\in[0,1]$ and $\mathcal{D}(\tilde{\Bb},\Bb)=0$ or $1$ if and only if the spaces spanned by $\Qb$ and $\tilde{\Qb}$ are the same or orthogonal.
In the following we introduce the clustering-related indexes to evaluate the recovery of group memberships: the Rand index, aRand index, Jacarrd index and Purity index. These indexes are commonly used for evaluating the performance of clustering methods, as adopted in \cite{10.1111/biom.13753} and \cite{tu2023} and the better the clustering performance, the closer these indexes approach 1. Besides, we resport $\hat{K}_{mean}$ as the average of the estimated group numbers. We also denote ``Freq ($a|b$)" as the number of underestimates (denoted by $a$) and overestimates (denoted by $b$) for the number of groups.

The simulation results of Scenario 1 and Scenario 2 are shown in the Table \ref{table2} and Table \ref{table3}. Firstly, when the noise level is low, i.e.,  $\kappa = 0.5 $, our method can correctly estimate the true number of groups and the grouping structures in all settings, while the TW method performs no better than ours. Secondly, as the noise level increases, both the methods tend to underestimate or overestimate the true number of groups, and the clustering performance correspondingly deteriorates. However, compared with the TW method, our method still shows a significant advantage under all settings. Notably, when  $T = 100$, $N = 150$ and $\kappa=1$, the TW method overestimates the true number of clusters 20 times, whereas our method only overestimates it once. Moreover, the performance of our proposed method in terms of clustering metrics, as well as the distance $\mathcal{D}(\tilde{\Bb},\Bb)$ and MSEs also show significant improvement in comparison with  those by the TW method. This is consistent with the conclusions derived in the last section, that our method is relatively more stable and shows more pronounced advantages when the noise level is high. In summary, for Scenario 1, in all cases, our method consistently outperforms the TW method in terms of all the indexes. The same conclusions can be drawn for  Scenario 2 by the results shown in Table \ref{table3}. 
\section{Real data analysis}\label{sec:5}
In this section, we apply the proposed method to identify the group structure of the daily returns from $N=49$ broadly diversified portfolios, ranging from January 4, 2016 to September 28, 2018. The dataset is widely studied using factor models and can be publicly downloaded from Kenneth R. French's webpage at \url{https://mba.tuck.dartmouth.edu/pages/faculty/ken.french/data_library.html}. In particular, \cite{guo2022homogeneity} and \cite{tu2023} ever explored the group structures of this dataset under the framework of factor models.
Accordingly, our goal is to study the homogeneity structure by our proposed method. In this section, we keep the stock returns from December 5, 2016 to November 30, 2017 as the training dataset to identify the group memberships, and the remaining test set is used for out-of-sample prediction. After preprocessing and standardization, the dimension of the training datasets returns as $(T,N)=(250,49)$.

\begin{figure}[htbp]
    \centering
    \includegraphics[width=17cm,height=9cm]{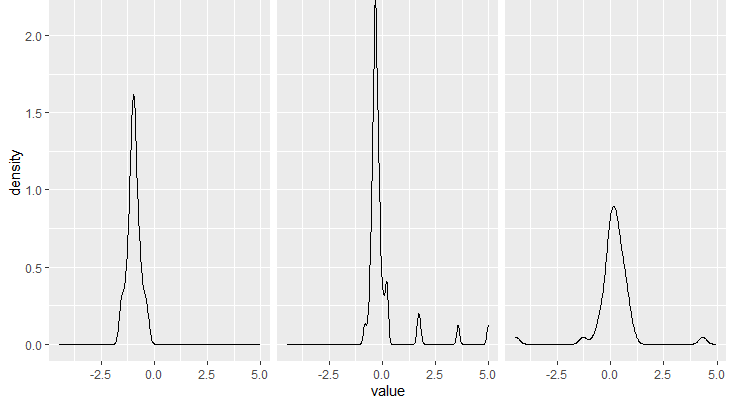}
    \caption{Kernel density functions of each columns of the estimated factor loading matrix  by the PCA method.}
    \label{figure1}
    \end{figure}

In the training set, we first determine the number of factors to be 3 by the information criterion $IC_2$ of \cite{bai2002determining}. To explore whether the dataset exhibits a grouping structure, we ignore the latent group structure and apply the \texttt{R} function \texttt{ggplot} to show kernel density function of factor loadings estimated by PCA estimators in Figure \ref{figure1}, from which we conclude that the distribution exhibits multiple modes with multiple prominent peaks, indicating the presence of a grouping structure in factor loadings. Hence, it is more reasonable to consider the proposed PPCA method for this real dataset.

\begin{figure}[!h]
    \centering
    \clearpage
\includegraphics[width=17cm,height=10cm]{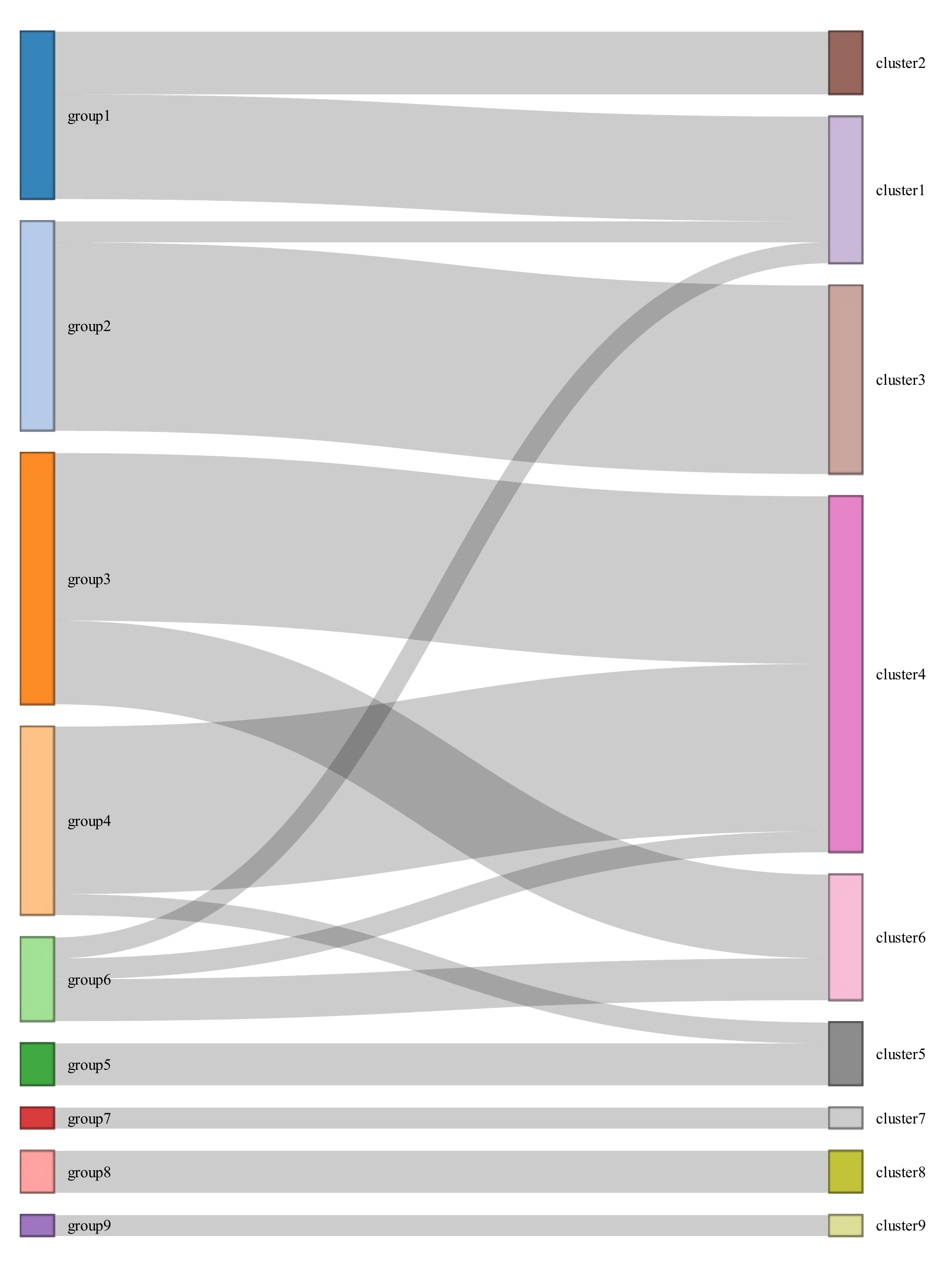}
    \caption{The Sankey diagram: the flows of portfolios from our identified groups (left) to those of \cite{tu2023} (right).}
    \label{figure3}
\end{figure}

Next, we investigate the homogeneity structure of factor loadings by our method and the TW method in \cite{tu2023}. The information criterion introduced in (\ref{equ:IC}) identifies 9 distinct groups by the PPCA estimators. For brevity, the estimated membership for all 9 groups with PPCA and PCA as initial values are reported in the supplement. To better illustrate the difference of the estimated group structures, we draw the Sankey diagram in Figure \ref{figure3} to show the flow of the group memberships estimated by the two methods. Clearly, our identified group structure has similarities to that identified by \cite{tu2023}. For example, both methods cluster the gold industry and coal industry into separate groups, while also grouping the mining and oil sectors together. However, there also exist significant distinctions between the group structures identified by our method and the TW method by \cite{tu2023}, one can refer to Table \ref{table6} and \ref{table7} in the appendix for more details.

Finally, we compare the out-of-sample prediction performance of our proposed method with that of \cite{tu2023}. For the accuracy measure, we follow \cite{guo2022homogeneity} to use the out-of-sample prediction error (OSPE) of a given month, which is defined as
$$\text{OSPE}=\frac{\sum_{i=1}^{49}\sum_{t\in \tau}(x_{ti}-\hat{x}_{ti})^2}{49\times |\tau|},$$
where $\tau$ denotes the set of days in this month remained after removing the missing values and $\hat{x}_{ti}$ is the predicted value of the $i$-th series at time $t$-th day of a month. Specifically, to forecast the daily returns in December 2017, we use a training set from December 2016 to November 2017 to estimate the loading matrix and then form the forecast with factors estimated from the full sample prior to this month. Next, to forecast the returns in January 2018, the training set is further augmented with data in December 2017. We continue this process until September 2018, and the OSPEs are sequentially calculated, presented in Table \ref{table8}, where the OSPE1 represents the OPSE results of the method by \cite{tu2023}, while the OPSE2 represents those by our proposed method. From Table \ref{table8}, we can see that our proposed method always outperforms   \cite{tu2023}'s method in terms of OPSEs.
\begin{table}[htbp]
\caption{The results of OPSEs. }
\renewcommand{\arraystretch}{1.3}
\centering
\label{table8}
\scalebox{0.9}{
\begin{tabular}{ccccccccccc}
	\toprule[2pt]
	&2017.12 & 2018.01 & 2018.02 & 2018.03& 2018.04&2018.05&2018.06&2018.07&2018.08&2018.09\\
	\hline
	OPSE1&1.8036&1.5782&3.4238&2.7635&2.3107&1.3026&2.0564&1.3511&1.5763&1.4675\\
	\hline
	OPSE2&1.7761&1.4815&3.4121&2.7463&2.1952&1.2439&1.9313&1.3066&1.4938&1.3973\\
	\bottomrule[2pt]
\end{tabular}
}
\end{table}

\section{Conclusions and discussions}\label{sec:6}
In this paper, we first propose a PPCA method with $\ell_2$-norm penalty for group pursuit under factor model and provide a closed-form solution. We give the asymptotic properties of the PPCA estimators. We further present an AHC approach for estimating the latent homogeneity structure in the loadings. Then we adopt the constrained least squares method to obtain the group-specific loading vectors. We also propose an information criterion to determine the unknown group number. Consistency and asymptotic normality are derived for both the PPCA estimators and the post-grouping estimators under very mild conditions. Moreover, we study the clustering and group number consistency under mild condition of minimum signal strength and group sizes. It is demonstrated that our proposed PPCA method outperforms the PCA method when latent groups exist. We also conclude that the PPCA estimators act as more promising initial values for subsequent clustering procedure. The real data analysis illustrates the practical merits of our methodology.

The current work can be extended in several directions. Firstly, the proposed PPCA method can be used to locate change points in the presence of group structure \citep{baltagi2017identification}. Secondly, the matrix factor model \citep{Han2020Tensor,Han2021Rank,chen2022factor,Yu2021Projected,He2023Vector,He2024Matrix,he2024online} has been well studied in the literature in the last few years and pursuit for two-way group homogeneity is  an interesting and challenging problem.
Such extensions deserve separate study and are left as our future work.

\section*{Acknowledgements}
The authors gratefully acknowledge  National Science Foundation of China (12171282),  Qilu Young Scholars Program of Shandong University.

\bibliographystyle{model2-names}
\bibliography{ref}

\end{document}